# Cross-layer Application-aware Power/Energy Management for Extreme Scale Science

[Position paper]

Ivan Rodero and Manish Parashar

Rutgers Discovery Informatics Institute and
NSF Cloud and Autonomic Computing Center,
Department of Electrical and Computer Engineering
Rutgers University
Contact: *{irodero, parashar}@rutgers.edu*

## 1. Motivation

High Performance Computing (HPC) has evolved over the past decades into increasingly complex and powerful systems. Current HPC systems consume several MWs of power, enough to power small towns, and are in fact soon approaching the limits of the power available to them. Estimates are with the given current technology, achieving exascale will require hundreds of MW, which is not feasible from multiple perspectives. Architecture and technology researchers are aggressively addressing this; however as past history is shown, innovation at these levels are not sufficient and have to be accompanied with innovations at higher levels (algorithms, programming, runtime, OS) to achieve the multiple orders of magnitude reduction – i.e., a comprehensive cross-layer and application-aware strategy is required. Furthermore, energy/power-efficiency has to be addressed in combination with quality of solutions, performance and reliability and other objectives and appropriate tradeoffs are required.

## 2. Challenges addressed

The main goal is to expand runtime energy/power management at multiple levels and how address this more effectively in a cross-layer manner and tradeoff energy/power with the quality of the solution. Current approaches try to optimize energy efficiency at different levels, such as, runtime or component-based power management; however, the optimization goals of these approaches may be conflicting. Also, current data management approaches typically operate on centralized data repositories and cannot handle the extreme rates of data generation and distribution scales. In these thrusts, we will explore autonomic mechanisms to manage energy efficiency through dynamic cross-layer adaptations in an integrated and holistic manner as illustrated in Figure 1.

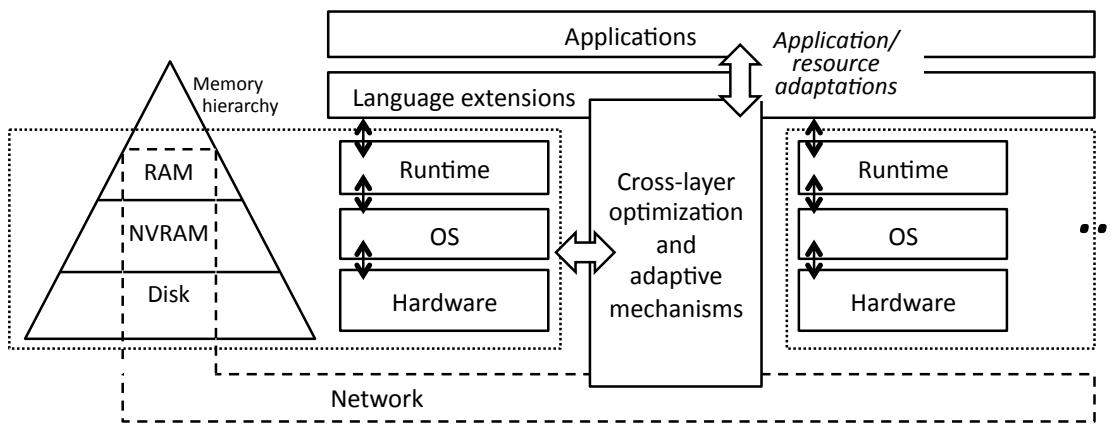

**Figure 1**: Conceptual view of the proposed cross-layer approach

The main challenges of such a cross-layer approach can be summarized as follows: (1) understanding the application patterns and how keeping the application in the loop can facilitate performing effective optimizations and adaptations at different levels in a cross-layer approach; (2) studying the implications of these adaptations to other dimensions such as performance, data analysis/management, and methods to effectively use all levels of the memory hierarchy [1]; and (3) understanding the control plane and the mechanisms that can be used to implement and orchestrate such a management approach.

## 3. Maturity

Our vision is revolving on a cross-layer approach to perform effective optimizations/adaptations. We have studied and modeled performance/energy tradeoffs of power management techniques at different levels and designed and evaluated model-driven autonomic optimizations and adaptations, including application-centric aggressive power management of data center's resources for HPC workloads [6], multi-channel DRAM power management [2] and application-aware cross-layer power management for PGAS applications on many-core platforms [3]. We have also explored energy-efficient and thermal-aware autonomic management of HPC workloads [5, 7], and created innovative testbeds to conduct experimental evaluation and explored cutting-edge platforms such as the Intel SCC many-core system.

## 4. Uniqueness

The proposed approach address energy efficiency keeping the application/workload in the loop, which facilitates performing more effective optimizations and adaptations, for example, by using language extensions to improve runtime decisions. A key aspect is providing these adaptations along multiple dimensions, for example, adapting the resources to the applications (e.g., using low power modes to conform to a power budget) or adapting the application to the resources (e.g., reducing the precision of the computation or changing convergence values in order to adapt the accuracy/quality of the solution to guarantee completion time in a given time frame).

## 5. Novelty

The core idea of our approach is having a control plane along multiple levels including OS mechanisms and interfaces that can be used to optimize multiple tradeoffs and perform adaptations. They can be automatically orchestrated at runtime, in a cross-layer way to respond to the heterogeneity and dynamics, both of the applications and the infrastructure. Our approach is based on model-based power/energy management; however, in contrast to existing approaches, we propose end-to-end optimizations and optimizations that integrate the application with the runtime (e.g., using language extensions). An example of this approach is to reactively and proactively manage and optimize adaptive simulations (e.g., SAMR applications [8]) by detecting conditions under which the parameters affecting the application deviate from their acceptable behavior or operation (e.g., simulation performance may be degrade severely due to low available low-latency memory and increased data transfer over the network) and determining the appropriate application reconfiguration strategy and the resources required to repartition work (e.g., amount of memory of each level of the hierarchy) to optimize the simulation performance and energy efficiency.

## 6. Applicability

The overall cross-layer approach can be applied to other relevant areas, such as, data processing, transfer and management, and the design and usage modes of deep memory hierarchies, in addition to energy efficiency. The proposed approach could also influence the design or adoption of advanced hardware (e.g., exposing sufficient power controls and low power modes for different subsystems), how applications are developed and data analyzed, and how the scientists interact with applications data. An example of how cross-layer optimizations can be applied to data management is transferring data over the network only when analyzing data (generated during simulation) in dedicated nodes is more energy/performance efficient than doing the analysis in-situ where the data is generated. Energy/power optimizations can be enabled, for example, using low power modes in the dedicated nodes when the data is in-transit. At the same time, deciding when/where to move/analyze the data can be influenced by application constraints or the application can be adapted to the resources based on energy/performance tradeoffs (e.g., accuracy of the solution or frequency of data analysis within a power budget).

**7. Effort**

The effort to effectively explore the proposed approach can be decomposed into different stages: (1) characterize relevant compute- and data-intensive applications and HPC benchmarks and explore performance/power tradeoffs to define models, (2) develop mechanisms, strategies, application extensions and usage modes to implement model-driven optimizations/adaptations, (3) enable cross-layer interactions and integrate them with the runtime system, and (4) explore our approach at larger scale using simulation [4] considering non-standard hardware configurations, hardware and software co-design and how they impact the applications and system.